\documentclass[12pt]{iopart}
\usepackage{amsmath} 
\usepackage[x11names]{xcolor}
\usepackage{endnotes}
\usepackage{subcaption}

\usepackage{mathabx, wasysym, graphicx,mathtools}

\begin{document}
\title[]{Probing fast oscillating scalar dark matter with atoms and molecules}

\author{Dionysios Antypas$^{1}$, Oleg Tretiak$^{1}$, Ke Zhang$^{1}$, Antoine Garcon$^{1}$, Gilad Perez$^{2}$, Mikhail G. Kozlov$^{3,4}$, Stephan Schiller$^{5}$, and Dmitry Budker$^{1,6}$}

\address{$^1$Johannes Gutenberg-Universit{\"a}t Mainz, Mainz, Germany, GSI Helmholtzzentrum für Schwerionenforschung, Darmstadt, Germany, Helmholtz-Institut Mainz, Mainz, Germany\\

$^2$Department of Particle Physics and Astrophysics,
Weizmann Institute of Science, Rehovot, Israel 7610001\\
$^3$Petersburg Nuclear Physics Institute of NRC ``Kurchatov Institute'', Gatchina 188300, Russia\\

$^4$St.~Petersburg Electrotechnical University
``LETI'', Prof. Popov Str. 5, 197376 St.~Petersburg, Russia\\

$^5$Institut für Experimentalphysik, Heinrich-Heine-Universität Düsseldorf, 40225 Düsseldorf,
Germany\\

$^6$Department of Physics, University of California, Berkeley, California 94720, USA\\

}

\ead{dantypas@uni-mainz.de}

\begin{abstract}
Light scalar Dark Matter with scalar couplings to matter is expected within several scenarios to induce variations in the fundamental constants of nature. Such variations can be searched for, among other ways, via atomic spectroscopy. Sensitive atomic observables  arise primarily due to possible changes in the fine-structure constant or the electron mass.  Most of the searches to date have focused on slow variations of the constants (i.e. modulation frequencies $<$ 1 Hz). In a recent experiment \mbox{[Phys. Rev. Lett. 123, 141102 (2019)]} called WReSL (Weekend Relaxion-Search Laboratory), we reported on a direct search for rapid variations in the radio-frequency band. Such a search is particularly motivated within a class of relaxion Dark Matter models. We discuss the WReSL experiment, report on progress towards improved measurements of rapid fundamental constant variations, and discuss the planned extension of the work to molecules, in which rapid variations of the nuclear mass can be sensitively searched for. 
\end{abstract}

%
\vspace{2pc}
\noindent{\it Keywords}: Dark matter, fundamental constant variations, relaxions, atomic spectroscopy, molecular spectroscopy.

%
%
%
%

\section{Introduction}

Direct astrophysical observations point to the existence of Dark Matter (DM), which is estimated to account for $\approx$80$\%$ of the total matter in the Universe \cite{PhysRevD.79.023519}.  This form of matter interacts with Standard Model (SM) matter gravitationally, but has  feeble (if any) other interactions, making efforts to uncover its origin, composition and properties challenging. The prevalent DM candidate scenario assumes that DM consists of Weakly Interacting Massive Particles (WIMPs), with mass in the 1-1000 GeV range. However searches with accelerators and scintillators  have not yielded a clear discovery yet \cite{Nature562(2020)}.
Within another class of motivated scenarios, DM consists of light bosonic particles (of mass $m_{\phi}$ in the $10^{-22}-1$ eV range), which form a classical field coherently oscillating at the Compton frequency of the underlying particle ($\omega_C=m_{\phi}$ \cite{Note1}). Several leading candidates, such as the QCD Axion, Axion-Like Particles (ALPs) and others, classified according to their spin, type of interaction(s) with SM particles and resulting physical observable \cite{PhysRevD.93.075029}, are the focus of a number of completed, ongoing or planned experiments \cite{RevModPhys.90.025008}. 

Within this light bosonic DM landscape,  attention is given to the possibility that DM couples to matter inducing oscillations in the fundamental constants (FC) of nature. Such FC oscillations are expected in cases where DM consists of particles emerging in higher-dimension theories (such as dilatons) with scalar coupling to the SM particles \cite{PhysRevD.91.015015, annurev-nucl-102014-022120} or relaxions, i.e. ALPs originally introduced to provide a solution to the hierarchy problem \cite{PhysRevLett.115.221801}. Within minimal models, the relaxion  can account for the observed DM in the Universe \cite{PhysRevD.100.115026}. The related phenomenology arises from a scalar coupling to SM matter via the relaxion-Higgs mixing \cite{JHEP06(2017)050,JHEP2020153}.

Motivated by a number of beyond-SM scenarios, a series of studies involving astronomical observations and laboratory experiments have focused on probing slow drifts in the FC (see, for example, \cite{RevModPhys.90.025008, ComptesRP16}, and references therein). The possibility that light scalar DM with coupling to matter induces FC oscillations, such as for example, oscillations in the fine structure constant $\alpha$, the electron mass $m_e$ or the quantum chromodynamics scale parameter $\Lambda_{\rm{QCD}}$, has motivated searches for such effects as well.   

A variety of approaches have been proposed to search for light scalar DM-induced oscillations in the FC. These include use of atomic  \cite{PhysRevD.91.015015, PhysRevLett.114.161301,PhysRevA.93.063630,PhysRevLett.120.173001, Dzuba2018}, or nuclear \cite{JHEP2020153} clocks to probe variations of $\alpha$ and detection of oscillations in $\alpha$, $m_e$ via laser interferometry \cite{PhysRevLett.114.161301,PhysRevA.93.063630}, comparison of optical cavities \cite{PhysRevLett.123.031304} or  resonant mass detectors \cite{PhysRevLett.116.031102}. Effects of Equivalence-Principle (EP)-violating forces  arising due to light scalar DM and means to detect those were considered in \cite{PhysRevD.93.075029, PhysRevD.98.064051}.  Analysis of astrophysical data from the early Universe has provided constraints on scalar DM-SM matter couplings \cite{PhysRevLett.115.201301}. Constraints on such interactions have also been provided via several laboratory results, such as: long-term comparison of Cs and Rb microwave clocks \cite{PhysRevLett.117.061301,PhysRevA.94.022111}, atomic spectroscopy in Dy \cite{PhysRevLett.115.011802}, comparison of an optical cavity with itself at different times \cite{DAMNED}, a comparison of an ultra-stable Si cavity with a Sr optical clock and H maser \cite{JYeSiCavity}, a terrestrial network of optical clocks \cite{Wcisoeaau4869}, EP and Fifth-Force (FF) apparatus \cite{PhysRevLett.119.231101,ProgNucPartPhys2009,PhysRevLett.120.141101,Wagner_2012}. 

While the most stringent limits on scalar DM-SM interactions in the low-frequency limit ($f<1$ Hz) come from atomic probes, at higher frequencies, those interactions have been better constrained by EP and FF experiments (see, for example, constraints presented in recent work \cite{JYeSiCavity}). This is because the former constraints are parametric in the FC oscillation frequency (as is shown in Section 2.1), while the latter have less pronounced dependence on frequency up to the $\approx$ 1 GHz cut-off of their sensitivity. This situation is different considering scenarios in which the terrestial laboratory is immersed in a halo of DM particles. This is conceivable for relaxions, which, as ALPs, can form Earth-bound or Solar-bound halos, leading to a local DM overdensity and in turn to an enhanced observability of this scenario \cite{CommunPhys(2020)}. Within such a scenario, the extension of direct searches with atomic probes is particularly motivated in the radio-frequency (rf) range. In this range, it is in fact legitimate to consider oscillations of dimensionful constants, which is not meaningful in the limit of low frequencies \cite{ADP2020}. 

In recent work we reported results of an atomic spectroscopy search for rapidly oscillating FC \cite{PhysRevLett.123.141102}. In the experiment, called the Weekend Relaxion-Search Laboratory (WReSL), the  optical transition frequency in atomic cesium (Cs) is compared to the frequency of a laser resonator with different sensitivity to FC variations. The resulting DM-related constraints on $\alpha$ and $m_e$ are competitive with those of EP experiments within part of the explored frequency range (20 kHz -- 100 MHz). Here we review the WReSL experiment, discuss experimental progress towards improved searches of rapid FC oscillations with WReSL, and introduce an extension of the technique to molecules, which will allow probing rapid variations of the nucleon mass and thus the $\Lambda_{\rm{QCD}}$ scale parameter. By `rapid', we refer to oscillations at frequencies $>$ 1 Hz. We stress however that the work and analysis presented here are equally relevant for both rapid and slow oscillations of the FC.

\section{\label{sec:section2}Probing light scalar Dark Matter with WReSL}

\subsection{DM-induced oscillations in fundamental constants and detection with WReSL}

Let us illustrate the atomic effects arising from a coupling of light scalar DM to SM particles, and how the WReSL experiment allows to look for observables emerging from rapidly oscillating FC.  In the work reported in \cite{PhysRevLett.123.141102} the possibly oscillating constants were $\alpha$ and $m_{\rm{e}}$, but the envisioned extension to molecules will focus on the nucleon mass. If DM consists of light bosons of mass $m_{\rm{\phi}}$ the corresponding DM field  oscillates coherently, according to

\begin{equation}
\label{eq:phi}
\phi(\vec{r},t)\approx\frac{\sqrt{2\rho_{\rm DM}}}{m_{\phi}}\sin({m_{\phi}t}),
\end{equation}

\noindent where $\rho_{\rm DM}\approx$ 0.4 GeV/cm$^2$ is the local DM density \cite{Catena_2010}. The oscillation is coherent within time scale associated with the quality factor $\omega/\Delta\omega\approx 2\pi/v_{\rm DM}^2\approx6\cdot10^6$, where $v_{\rm DM}\approx 10^{-3}$ is the virial velocity of the DM field \cite{PhysRevLett.55.1797} (coherence is longer within relaxion-halo scenarios \cite{CommunPhys(2020)}). In the case of dilatonic or relaxion DM the constants $\alpha$ and $m_{\rm{e}}$ acquire a component which oscillates at the frequency $m_{\rm{\phi}}$:

\begin{equation}
\label{eq:alphaVar}
\alpha(\vec{r},t)=\alpha_0\big[1+g_{\gamma}\phi(\vec{r},t)\big],
\end{equation}
\begin{equation}
\label{eq:mVar}
m_{\rm{e}}(\vec{r},t)=m_{\rm{e,0}}\Big[1+\frac{g_{\rm{e}}}{m_{\rm{e,0}}}\phi(\vec{r},t)\Big].
\end{equation}

\noindent The parameters $\alpha_{\rm{0}}$ and $m_{\rm{e,0}}$ refer to the time-averaged values of the constants, and $g_{\rm{\gamma}}$, $g_{\rm{\e}}$ are DM couplings to the photon and the electron, respectively, that the experiment allows to investigate. This investigation exploits the modulation of the frequency of an atomic transition arising from oscillations in $\alpha$  or $m_{\rm{e}}$. As the atomic levels have energy proportional to the Rydberg constant $R_{\infty}=(1/2)m_e\alpha^2$, the frequency of an optical transition $f_{\rm{at}}$ acquires an oscillatory component. The resulting fractional modulation in $f_{\rm{at}}$ is given by

\begin{equation}
\label{eq:deltafatom}
\frac{\delta f_{\rm at}}{f_{\rm at}}=
2\frac{\delta\alpha}{\alpha_0}+
\frac{\delta m_{\rm e}}{m_{\rm e,0}}
=\Big(2g_{\rm{\gamma}}+\frac{g_{\rm{e}}}{m_{\rm{e,0}}}\Big)\frac{\sqrt{2\rho_{\rm DM}}}{m_{\rm{\phi}}}.
\end{equation}

The WReSL experiment employs optical spectroscopy of an atomic vapor to probe rapid variations in $f_{\rm{at}}$. A laser is tuned in frequency to resonantly excite atoms; the experimental signal contains information about  deviations between the laser frequency $f_{\rm{L}}$ and $f_{\rm{at}}$,  i.e. the difference $f_{\rm{at}} - f_{\rm L}$, where the two frequencies are close to each other (on the optical scale). In the absence of identifiable noise sources, this difference is to be attributed to variations of FC and hence, to a detection of the couplings $g_{\rm{\gamma}}$ and/or $g_{\rm{\e}}$.  However, the sensitivity in detection of $g_{\rm{\gamma}}$ and $g_{\rm{\e}}$ from a comparison between $f_{\rm{L}}$ and $f_{\rm{at}}$ is not the same over the entire frequency range over which the search is performed \mbox{(20\,kHz$-$100\,MHz)}
\cite{Note2}. One has to consider the following two effects, which are discussed in detail in \cite{ADP2020,PhysRevLett.123.141102}: i) the sensitivity of the laser frequency itself to oscillations in $\alpha$ and $m_{\rm{e}}$, and ii) the response of the atomic signal to oscillations of the constants, at frequencies comparable to, or larger than the linewidth $\Gamma/2\pi$ of the optical transition.  As $f_{\rm{L}}$ is inversely proportional to the length $L$ of the laser resonator, which scales as $L\propto$ $1/\alpha m_{\rm{e}}$,  there is in fact modulation in $f_{\rm{L}}$ with fractional amplitude 

\begin{equation}
\label{eq:deltafL}
\frac{\delta f_{\rm L}}{f_{\rm{L}}}=
\frac{\delta\alpha}{\alpha_0}+
\frac{\delta m_{\rm e}}{m_{\rm{e,0}}}=
\Big(
g_{\rm{\gamma}}+\frac{g_{\rm e}}{m_{\rm e,0}}
\Big)
\frac{\sqrt{2\rho_{\rm DM}}}{m_{\phi}}.
\end{equation}

\noindent This modulation occurs at frequencies lower than the cutoff frequency $f_{\rm{c1}}$ for sound propagation in body of the laser resonator, which for the WReSL laser system is $f_{\rm{c1}}\approx$ 50 kHz \cite{PhysRevLett.123.141102}. A comparison of Eq. (\ref{eq:deltafatom}) and (\ref{eq:deltafL}) shows that for $f<f_{\rm{c1}}$ there is reduced sensitivity to oscillations of $\alpha$ and there is no sensitivity to oscillations of  $m_{\rm{e}}$. Another cutoff is imposed by the decay of the atomic response at frequencies larger than the  linewidth $\Gamma/2\pi$. The sensitivity  above this cutoff $f_{\rm{c2}}=\Gamma/2\pi$ (on order of MHz) decays as  $1/f$. To summarize, the relation between the measured fractional modulation $\delta (f_{\rm{at}}-f_{\rm{L}})/f_{\rm{at}}$ and the couplings $g_{\rm{\gamma}}$, $g_{\rm{\e}}$ is assumed as follows

\begin{equation}
\label{eq:fatcases}
\frac{\delta (f_{\rm{at}}-f_{\rm{L}})}{f_{\rm{at}}}=\begin{dcases}
       g_{\rm{\gamma}}\frac{\sqrt{2\rho_{\rm DM}}}{m_{\rm{\phi}}}h_{\rm{at}}(f), & f\leq f_{\rm{c1}} \;\;\;\;\;\;\;\;\;\;\;\;\;\;\; \\
              \Big(2g_{\rm{\gamma}}+\frac{g_{\rm{e}}}{m_{\rm{e,0}}}\Big)\frac{\sqrt{2\rho_{\rm DM}}}{m_{\rm{\phi}}}h_{\rm{at}}(f), & f> f_{\rm{c1}}. \\
    \end{dcases}
\end{equation}

\noindent The function $h_{\rm{at}}(f)$ is the atomic response function, with $h_{\rm{at}}(f)\approx1$ for $f\ll\Gamma\ /2\pi$ and $h_{\rm{at}}(f)\rightarrow \Gamma/ (2\pi f)$ for $f\gg\Gamma/2\pi$.  The function $h_{\rm{at}}(f)$ can be determined by checking the response of the atoms to a known frequency modulation imposed on the laser field exciting them. From an experimental point of view, it is important to maximize the apparatus fractional frequency sensitivity to obtain optimal sensitivity in detection of the couplings $g_{\rm{\gamma}},g_{\rm{e}}$. 


\subsection{The WReSL experiment and preliminary results}

\begin{figure}

\;\;\;\;\;\;\;\;\;\;\;\;\;\;\;\;\;\;\;\;\;\;\;\;
\includegraphics[width=10cm]{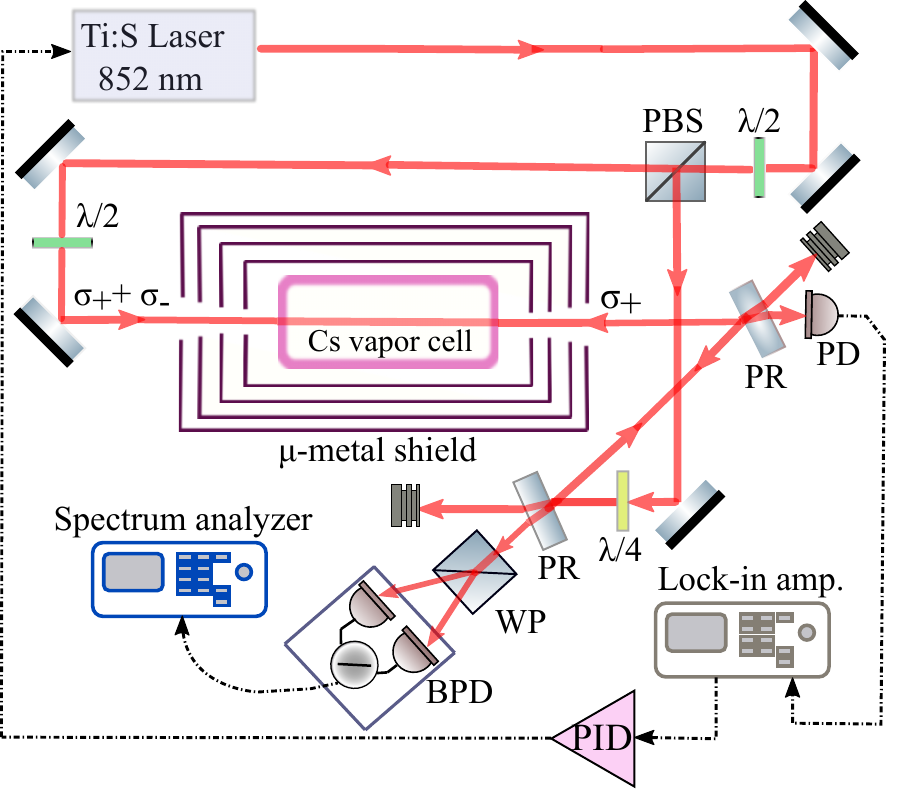}
\caption{Simplified schematic of the polarization- spectroscopy setup employed in the 2019 WReSL run \cite{PhysRevLett.123.141102}. PBS: polarizing beamsplitter, $\lambda$/2: half-wave plate, $\lambda$/4: quarter-wave plate, PR: partial reflector, WP: Wollaston prism, PD: photodetector, BPD: balanced photodetector. Adapted from \cite{PhysRevLett.123.141102}.}
\label{fig:apparatus}
\end{figure}

\begin{figure}

\;\;\;\;\;\;\;\;\;\;\;\;\;\;\;\;\;\;\;\;\;\;\;\;
\includegraphics[width=10cm]{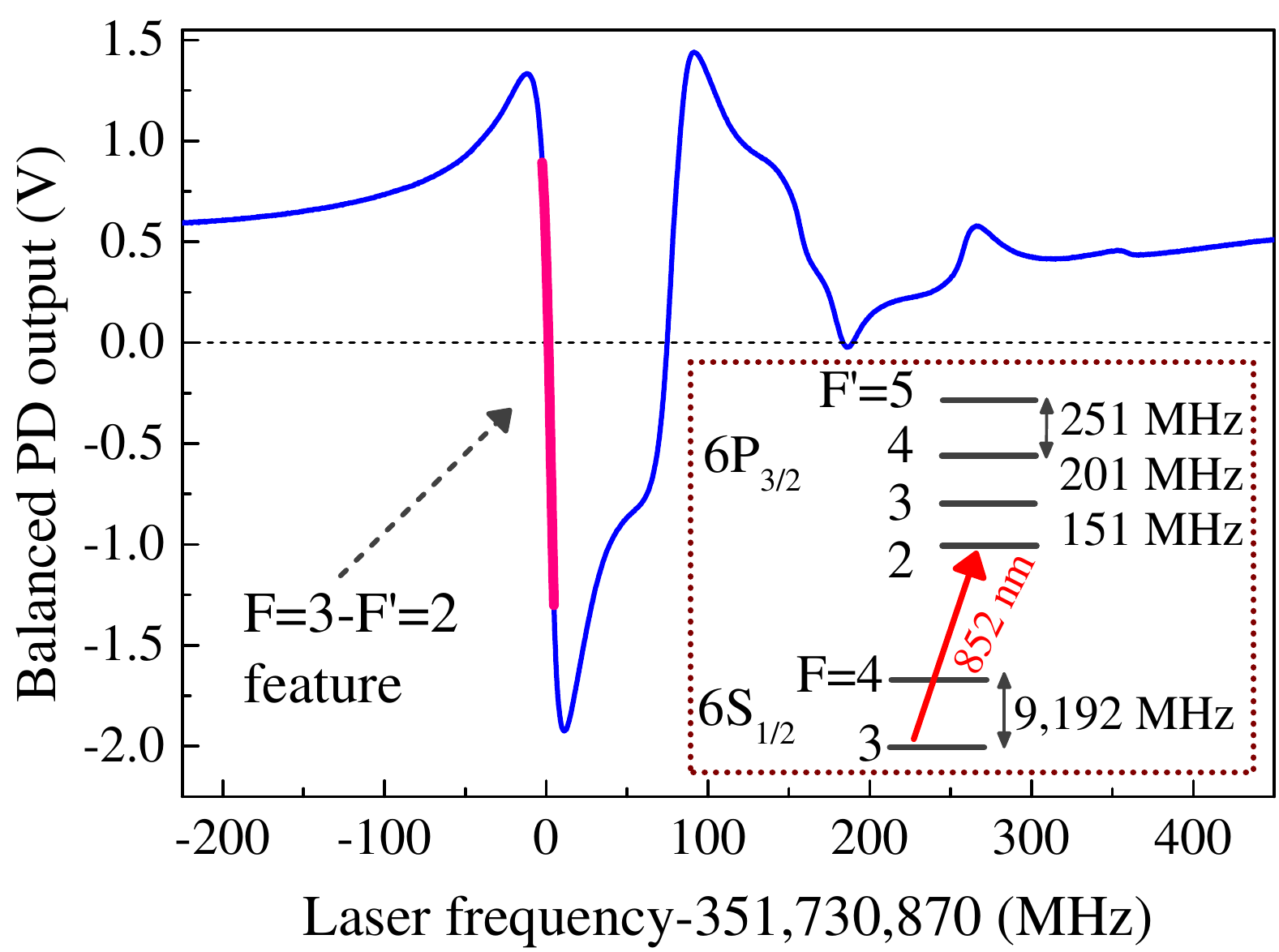}
\caption{ Polarization spectroscopy on the 6S$_{1/2}$ $F=3\rightarrow $ 6P$_{3/2}$ $F'=2,3,4$ transitions obtained with the apparatus used in the 2019 WReSL run \cite{PhysRevLett.123.141102}. The pink line indicates the feature employed for frequency discrimination. The measured slope is used to extract constraints on the quantity $\delta (f_{\rm{at}}-f_{\rm{L}})/f_{\rm{at}}$.  Shown in the inset are the hyperfine levels of the ground 6S$_{1/2}$ state and excited 6P$_{3/2}$ state. Adapted from \cite{PhysRevLett.123.141102}. }

\label{fig:spectrum}
\end{figure}

A preliminary WReSL run was carried out in 2019, whose results on scalar DM constraints were reported in \cite{PhysRevLett.123.141102}. Optical spectroscopy was done on the D2 line of cesium (Cs) vapor to search for rapid FC oscillations. A schematic of the setup is shown in Fig.\,\ref{fig:apparatus}. The apparatus implements polarization spectroscopy \cite{Demtroder} on the \mbox{6S$_{1/2}$ $\rightarrow $ 6P$_{3/2}$} transition (natural linewidth $\Gamma/2\pi\approx5.2$ MHz), which is excited with light from a Ti:Sapphire laser. A benefit of this method is that the (nearly Doppler-free) resonances  obtained for each of the hyperfine components of the transition have a dispersion-shape profile, that can be conveniently employed as a frequency discriminator ({Fig. \ref{fig:spectrum}}). With the laser frequency tuned to the zero-crossing of the lineshape, spectral analysis of the polarization spectroscopy signal is done to detect amplitude modulation which is expected in the presence of DM-induced oscillations in the transition frequency $f_{\rm{at}}$ and laser frequency $f_{\rm{L}}$. In the absence of detection of such oscillations, the measured power noise spectrum of the signal is used to constrain the fractional variation $\delta (f_{\rm{at}}-f_{\rm{L}})/f_{\rm{at}}$  within the investigated frequency range (20 kHz-100 MHz),  and through Eq.\,(\ref{eq:fatcases}) constrain the couplings $g_{\rm{\gamma}}$ and $g_{\rm{\e}}$. 

\begin{figure}[]
\centering
\begin{subfigure}{.49\textwidth}
  \centering
  \includegraphics[width=01\linewidth]{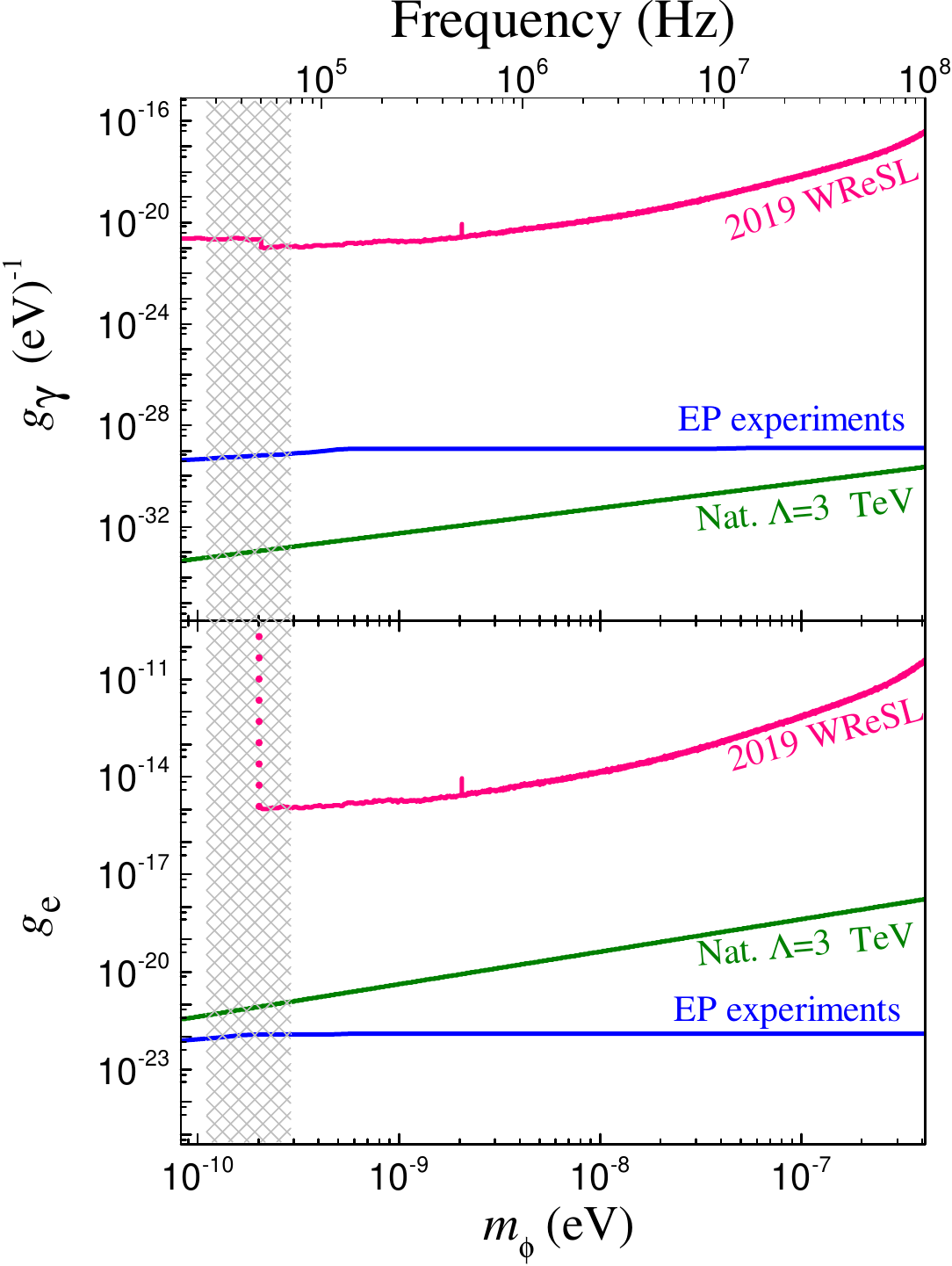}
\end{subfigure}
\begin{subfigure}{.49\textwidth}
  \centering
  \includegraphics[width=1\linewidth]{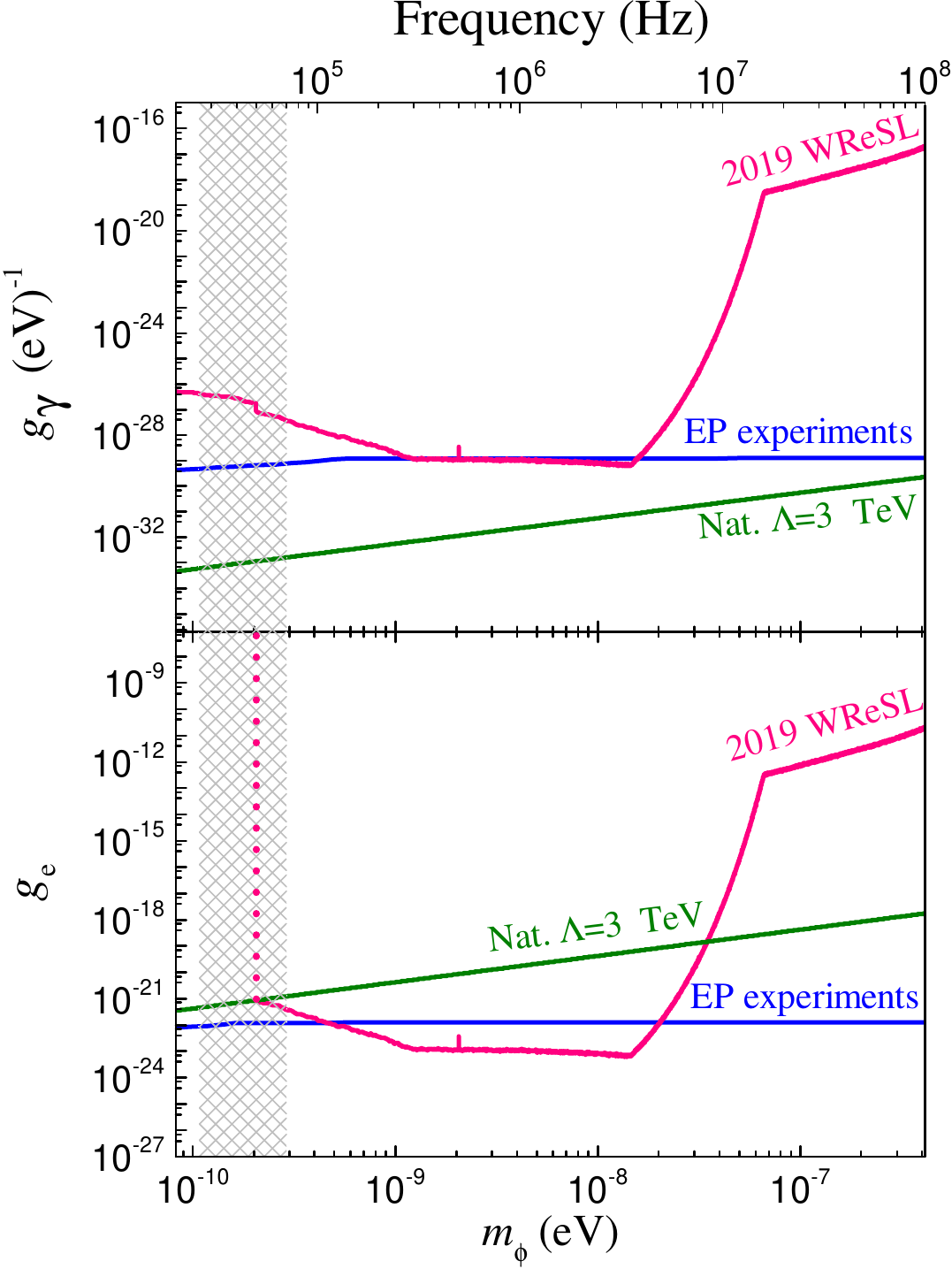}
\end{subfigure}
\caption{ Constraints in the couplings $g_{\gamma}$ and $g_e$  [shown at the 95\% confidence level (CL)], extracted from the earlier WReSL run \cite{PhysRevLett.123.141102}. Left: Limits considering the average galactic DM density. The constraint in green comes from the requirement to maintain Naturalness \cite{PhysRevLett.116.031102, PhysRevD.93.075029}. Right: Extracted limits (shown at the 95\% CL) within the scenario of a relaxion Earth halo. The shaded area in the plots indicates a region around the laser resonator cutoff frequency $f_{\rm{c1}}\approx$ 50 kHz, where careful consideration of the resonator response is needed to accurately determine the transition in sensitivity of probing $g_{\rm{\gamma}}$, $g_{\rm{e}}$. The limits from EP experiments are from \cite{SmithPRD1999,SchlammingerPRL2008}. Constraints of \cite{PhysRevLett.123.141102} in the range 40-100 MHz are scaled up here by factor in the range 1-2, to account for improved apparatus calibration. Adapted from \cite{PhysRevLett.123.141102}. }
\label{fig:constraints}
\end{figure}

   As no FC oscillations were detected in the first WReSL run \cite{PhysRevLett.123.141102},  limits on scalar DM couplings to the photon and the electron were placed from analysis of the spectrum of the $\delta (f_{\rm{at}}-f_{\rm{L}})/f_{\rm{at}}$ parameter [Eq. (\ref{eq:fatcases})], which was constrained to better than $10^{-14}$ in part of the explored frequency range. Extracted constraints are shown in Fig. (\ref{fig:constraints}). They are computed for the average galactic DM density $\rho_{\rm DM}=$ 0.4 GeV/cm$^3$ , but also within the scenario of a relaxion halo gravitationally bound by Earth \cite{CommunPhys(2020)}. In the latter case the DM field density $\rho_{\rm{DM}}$ is greatly enhanced.  The resulting DM overdensity that we consider here is calculated in \cite{CommunPhys(2020)}, and its effects on constraining scalar DM interactions are mostly pronounced in the $\approx$ 1 MHz range. This range has remained out of reach for most of the experimental searches for scalar DM, that with the exception of \cite{OzeriSrDD} have probed the range below 1 Hz. As seen in Fig.\,\ref{fig:constraints} the limits on $g_{\rm{\gamma}}$, $g_{\rm{e}}$ are tighter in the vicinity of $\approx$1 MHz, relative to those provided by EP experiments, in which there is no sensitivity enhancement with consideration of a DM halo scenario. 
   
 We note that the coupling considered in Eqs.\,\eqref{eq:alphaVar} and \eqref{eq:mVar} describes linear interactions between the scalar and the SM fields. As a result, the potential induced between two SM sources, that is bounded by experiments that search for violation of the EP, is induced by virtual exchange of the scalar particle and is independent of its background density \cite{Note4}. 

\section{Towards an improved search for scalar DM with WReSL }

Substantial technical improvements compared to the earlier WReSL setup \cite{PhysRevLett.123.141102} have yielded a greatly enhanced experimental sensitivity. The primary improvement has been a dramatic increase in the measurement duty cycle. The search for FC oscillations is done via spectral analysis of the polarization spectroscopy signal (Fig. \ref{fig:spectrum}). In the first WReSL run the analysis was carried out with a commercial, swept-frequency spectrum analyzer, which performed analysis in a sub-optimal way, resulting in an low effective duty cycle (of order $10^{-4}$), thus necessitating a long acquisition time for a given target sensitivity. Spectral analysis is now done with a homebuilt, computer-based Fast-Fourier-Transform (FFT) spectrometer, whose implementation is based on \cite{5634388}. The new spectrometer incorporates a 250 MSa/s data-acquisition card and an efficient graphics-processing unit, and performs real-time FFT analysis in our 20 kHz-100 MHz range of interest with $\approx$ 1 Hz resolution. This increased efficiency allows for continuous recording of the whole spectral range and brings the measurement duty cycle close to unity, yielding a sensitivity enhancement in detection of FC oscillations of order 100. 

Additional enhancement came from an increase in the power of the light exciting atoms, leading to improved photon shot-noise level in the measurement of the signal of Fig. \ref{fig:spectrum}. This required replacement of the vapor cell with one of larger size, allowing for  $\approx \times$4 larger-diameter laser beams traversing the cell, without significantly increasing saturation and broadening of the atomic-transition lineshape. The upgrade lead to a $\times 4$ sensitivity improvement. 

\begin{figure}
\;\;\;\;\;\;\;\;\;\;\;\;\;\;\;\;\;\;\;\;\;\;\;\;
\includegraphics[width=10cm]{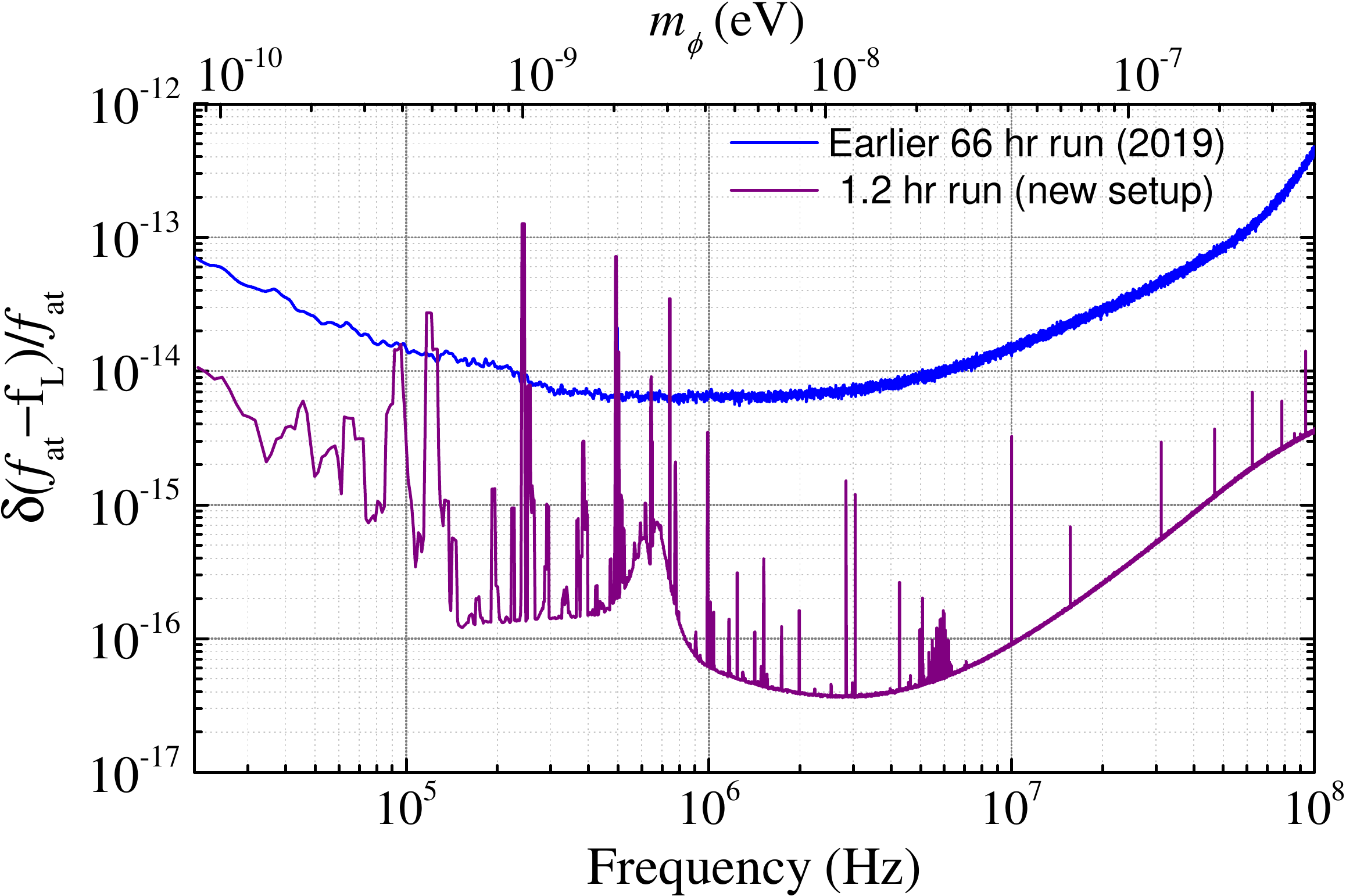}
\caption{  Comparison of constraints on $\delta (f_{\rm{at}}-f_{\rm{L}})/f_{\rm{at}}$ shown for the earlier 2019 run \cite{PhysRevLett.123.141102} and the new, ongoing experiment with improved sensitivity apparatus. Constraints are shown at the 95\% CL. The 2019 data are scaled up by factor in the range 1-2  between 40-100 MHz, to account for improved apparatus calibration.  A series of technical noise peaks in the new data are being investigated.}
\label{fig:dvv}
\end{figure}

Figure \ref{fig:dvv} shows an example of the measured spectrum of $\delta (f_{\rm{at}}-f_{\rm{L}})/f_{\rm{at}}$ with the upgraded apparatus. Data acquisition of duration $\approx$ 72 min results in up to $\approx$100  greater sensitivity compared to the previous results (also shown), for which data were acquired for 66 hrs. With longer integration we anticipate an additional factor in excess of $\times$10 in detection of $\delta (f_{\rm{at}}-f_{\rm{L}})/f_{\rm{at}}$ variations.  A complication in interpreting the spectrum arises due to the enhanced sensitivity; far more spurious peaks due to technical  noise sources appear, that must be properly investigated to exploit the full power of the method to detect, or constrain FC oscillations. There are various ways that some peaks can be rejected, for example, based on the fact that they appear in both signals, from the atoms and from the optical cavity, while the latter cannot respond to fast-oscillating DM field. Such peaks thus likely originate from frequency modulation of the laser light. Additionally, some peaks appear in multiplets with fixed separation (of exactly 20\,Hz). These are likely related to electromagnetic pick-up from some device, for example, a switching power supply in or near the building. Some of the peaks have linewidths inconsistent with models of galactic DM. 
This investigation is ongoing.

\section{Extension to molecules: probing oscillations of the nuclear mass }

While atomic systems are primarily sensitive to variations in $\alpha$ and $m_{\rm{e}}$,  molecules  are additionally sensitive to variations of the nuclear mass $M$. This can be exploited to search for DM-induced oscillations in $M$, or to leading order, in the $\Lambda_{\rm{QCD}}$ scale parameter, as this parameter predominantly determines the masses of protons and neutrons \cite{CFK2009}. Additional contributions can arise due to the couplings to the quarks. For a survey of the proposed, ongoing and completed activities involving searches for FC variations using molecules we refer the reader to \cite{RevModPhys.90.025008,Hanneke_2020}. We note that searches for DM with molecular spectroscopy that involve signatures other than FC variations, have also been proposed (see for example Refs. \cite{ArvanitakiPRX2018,FichetPRL2018}). 

In this section we analyse the molecular sensitivity to variations in $M$. This analysis is useful in guiding future molecule-based searches for FC oscillations, slow or rapid. In their presence, the nuclear mass can be expressed as 
\begin{equation}
\label{eq:MVar}
M(\vec{r},t)=M_{\rm{0}}\Big[1+\frac{g_{\rm{n}}}{M_{\rm{0}}}\phi(\vec{r},t)\Big],
\end{equation}

\noindent where $M_{\rm{0}}$ is the time-averaged mass, and $g_{\rm{n}}$ is the coupling of the DM field $\phi$ to the nucleons.

In molecules, in addition to the electronic levels, there are vibrational and rotational levels, with the total energy $E_{\rm{mol}}$ of the system given by 
\begin{equation}
\label{eq:Emol}
E_{\rm{mol}}=E_{\rm{el}}+E_{\rm{vib}}+E_{\rm{rot}}. 
\end{equation}

\noindent The terms  in Eq.\,(\ref{eq:Emol}) have generally different sensitivities to a change in the nuclear mass. To examine the sensitivity of these terms to changes in $M$ we will write them such that the dependence on $M$ is explicit. As we explain below, it is the vibrational 
 energy that 
is 
most sensitive to $M$ variations. 

\noindent The energy of a given electronic level $E_{\rm{el}}$ is proportional to the Rydberg constant $R_{\infty}$, so that $E_{\rm{el}}=
2C_{\rm{el}} h c R_{\infty}$, where $C_{\rm{el}}$ is a constant of order unity, independent of FC in the non-relativistic approximation. 
 In the electronic energy, $C_{\rm el}$  acquires dependence on $\alpha$  when relativistic effects are considered:
\begin{equation}
\label{eq:Cel}
C_{\rm el}=C_0+C_1(\alpha Z_{\rm eff})^2+.... 
\end{equation}
\noindent The coefficients $C_i$ are of order unity. In the electronic transitions considered here (accessible by conventional laser spectroscopy), it is the weakly bound electrons that transition between orbitals. For these, the effective nuclear charge $Z_{\rm eff}\simeq1$. Thus, for the purpose of this discussion, the dependence of $C_{\rm el}$ on FC can be neglected.

Going beyond the approximation of an infinite nuclear mass necessitates  replacing $m_{\rm{e}}$ with the reduced mass $m_{\rm{e}}M/(m_{\rm{e}}+M)$, so that 

\begin{equation}
\label{eq:Eelapprox}
E_{\rm{el}}=C_{\rm{el}}\frac{m_{\rm{e}}M}{m_{\rm{e}}+M}\alpha^2 c^2
\simeq C_{\rm{el}}m_{\rm{e}}\alpha^2 c^2\Big(1-\frac{m_{\rm{e}}}{M}\Big). 
\end{equation}

\noindent The electronic energy change $\delta E_{\rm{el}}^{(M)}$ with a change $\delta M$ is given by

\begin{equation}
    \label{eq:deltaEel}
         \delta E_{\rm{el}}^{(M)}=2C_{\rm{el}} h c R_\infty \Big(\frac{m_{\rm{e,0}}}{M_{\rm{0}}}\Big)\frac{\delta M}{M_{\rm{0}}}= E_{\rm{el}}\Big(\frac{m_{\rm{e,0}}}{M_{\rm{0}}}\Big)\frac{\delta M}{M_{\rm{0}}},
    \end{equation}

\noindent where the superscript ``$(M)$" denotes that only the sensitivity to $M$ is being considered. Note that this sensitivity estimate is accurate up to a factor of order unity.

The  vibrational energy is given by
\begin{equation}
\label{eq:Evib1}
E_{\rm{vib}}=\omega_{\rm{e}}\Big(\upsilon+\frac{1}{2}\Big) -\omega_{\rm{e}}\chi_{\rm{\e}}\Big(\upsilon+\frac{1}{2}\Big)^2,
\end{equation}

\noindent where $\omega_{\rm{e}}$ is the fundamental vibrational transition energy, $\chi_{\rm{\e}}$ is an anharmonicity constant, and anharmonic terms of higher order in the vibrational quantum number $\upsilon$ are omitted. 

\noindent According to the Born-Oppenheimer theory, the vibrational energy $\omega_{\rm{e}}$ is given by

\begin{equation}
\omega_{\rm{e}}=2 C_{\rm{}vib} h c R_\infty \Big(\frac{m_e}{M_{\rm{r}}}\Big)^{1/2},
\end{equation}
where $M_{\rm{r}}\propto M$ is the reduced nuclear mass of the molecule and $C_{\rm{vib}}$ is a constant of order unity.  
The lowest-order corrections to $C_{\rm{vib}}$ are proportional to $(\alpha Z_{\rm eff})^2$, similar to $C_{\rm{el}}$  in Eq.\,(\ref{eq:Cel}), due to relativistic effects. 
They may be neglected here.
 For simplicity, we neglect the dependence of the anharmonicity constant on FC (the constant $\chi_{\rm{e}}$ scales as $(m_{\rm{e}}/M)^{1/2}$ \cite{JBU2014}).

\noindent A change $\delta M$ induces a change in  $E_{\rm{vib}}$ 
%


\begin{equation}
\label{eq:deltaEvib2}
    \delta E_{\rm{vib}}^{(M)}=\frac{1}{2} E_{\rm{vib}} 
     \Big(-\frac{\delta M}{M_{\rm{0}}} \Big).
     \end{equation}

\noindent The rotational energy is given by
\begin{equation}
\label{eq:Erot1}
E_{\rm{rot}}=BJ(J+1),
\end{equation}

\noindent where $B$ is the rotational constant, $J$ is the rotational quantum number of the level, and higher-order contributions are omitted. The rotational constant is $\hbar^2/(2 I)$, where $I$ is the  moment of inertia. For a diatomic molecule, it takes the form $I=M_{\rm r}d^2$,
with $d$ being the distance between the nuclei. Since $d$ is proportional to the Bohr radius  with proportionality constant of order unity, we can reexpress  
the rotational energy as

\begin{equation}
\label{eq:Erot2}
E_{\rm{rot}}=C_{\rm{rot}} h c R_\infty \Big(\frac{m_e}{M_{\rm{r}}}\Big)J(J+1),
\end{equation}

\noindent where $C_{\rm{rot}}$ is a constant  of order unity. 

The dependence of $\delta E_{\rm{rot}}^{(M)}$ on $\delta M$ is




\begin{equation}
\label{eq:deltaErot2}
\delta E_{\rm{rot}}^{(M)}=E_{\rm{rot}} \Big(-\frac{\delta M}{M_{\rm{0}}}\Big).
\end{equation}

One can make use of eqs.\,\eqref{eq:deltaEel}, \eqref{eq:deltaEvib2}, and \eqref{eq:deltaErot2} to evaluate the contributions to the variation of a transition frequency $f_{\rm{mol}}=f_{\rm{el}}+f_{\rm{vib}}+f_{\rm{rot}}$, considering optical excitation from the ground to an excited electronic level, with  $(\nu,J)\rightarrow(\nu',J')$. The rotational quantum number selection rule for a one-photon, electric-dipole allowed electronic transition in a diatomic molecule is $\Delta J=J'-J= 0,\pm 1$. The respective 
frequency changes  $\delta f^{(M)}$ due to the nuclear mass only are:

\begin{equation}
    \label{eq:deltafEel1}
        h \delta f_{\rm el}^{(M)}=(E'_{\rm{el}}-E_{\rm{el}})\Big(\frac{m_{\rm{e,0}}}{M_{\rm{0}}}\Big)\frac{\delta M}{M_{\rm{0}}},
    \end{equation}

\begin{equation}
\label{eq:deltafvib}
     h \delta  f_{\rm vib}^{(M)}= \frac{1}{2} \bigg\{ \Big[\omega'_{\rm{e}}\Big(\upsilon'+\frac{1}{2}\Big) -(\omega_{\rm{e}}\chi_{\rm{\e}})'\Big(\upsilon'+\frac{1}{2}\Big)^2\Big]-\Big[\omega_{\rm{e}}\Big(\upsilon+\frac{1}{2}\Big) -\omega_{\rm{e}}\chi_{\rm{\e}}\Big(\upsilon+\frac{1}{2}\Big)^2\Big]\bigg\}\Big(-\frac{\delta M}{M_{\rm{0}}} \Big),
\end{equation}
and 

\begin{equation}
\label{eq:deltafrot}
h \delta  f_{\rm rot}^{(M)}=\big[B'J'(J'+1)-BJ(J+1)\big] \Big(-\frac{\delta M}{M_{\rm{0}}}\Big). 
\end{equation}

\noindent Let us estimate the relative scaling of the variations $\delta f_{\rm{el}}^{(M)}$, $\delta f_{\rm{vib}}^{(M)}$ and $\delta f_{\rm{rot}}^{(M)}$. We consider a specific system: iodine $^{127}$I$_2$, a homonuclear diatomic molecule with \mbox{$M\approx 127$ u}, in which electric-dipole transitions in the visible range between the ground $X\,^1\Sigma_g$ and excited $B\,^3\Pi_u$ electronic level are conveniently accessible. 
The 
constants related to the molecular energy are $E'_{\rm{el}}-E_{\rm{el}}=(h\,c)\,15769$\,cm$^{-1}$,
$\omega'_{\rm{e}}=(h \,c)$\,125.7\,cm$^{-1}$, $(\omega_{\rm{e}}\chi_{\rm{e}})'= (h\, c) \,0.764$\,cm$^{-1}$, $B'=(h \,c)\, 0.029$\,cm$^{-1}$, 
$\omega_{\rm{e}}=(h\, c)$\,214.5\,cm$^{-1}$, $\omega_{\rm{e}}\chi_{\rm{e}}=(h \,c)\, 0.615$\,cm$^{-1}$, 
$B=(h\, c)\, 0.0374$\,cm$^{-1}$ \cite{GLI2ATLAS}. 

Taking as an example a  transition at $\lambda=565.409$\,nm \cite{GLI2ATLAS} with $(\upsilon=1, J=80)\rightarrow(\upsilon'=21,J'=79)$  we obtain:


\begin{equation}
    \label{eq:deltafEel2}
        \delta f_{\rm el}^{(M)}\simeq2.0\,\hbox{\rm GHz}\,\frac{\delta M}{M_{\rm{0}}}\,,
    \end{equation}

\begin{equation}
    \label{eq:deltafvib2}
       \delta f_{\rm vib}^{(M)}\simeq-30\,\hbox{\rm THz}\,\frac{\delta M}{M_{\rm{0}}}\,,
    \end{equation}
    \noindent and
    
    \begin{equation}
    \label{eq:deltafrot2}
   \delta f_{\rm rot}^{(M)}\simeq1.8\,\hbox{\rm THz}\,\frac{\delta M}{M_{\rm{0}}}\,.
    \end{equation}

\noindent The relative scaling of the above quantities is $ \vert \delta f_{\rm{vib}}^{(M)}\vert: \vert  \delta f_{\rm{rot}}^{(M)}\vert: \vert  \delta f_{\rm{el}}^{(M)}\vert \sim$  
{\bf $1.5\times10^4$ : $9\times10^2$ : $1$}. We see that the vibrational part of this molecular transition is the most sensitive to $M$ variations, with a smaller effect in the rotational and the electronic levels.

Primarily because transitions involving large change $\Delta\upsilon$ are allowed whereas the change in $J$ follows the selection rule 
$\vert\Delta J\vert\le1$, it is the choice of vibrational levels 
that has to be made optimally for a high-sensitivity search for variations of $M$. There is in fact an optimal $\Delta\upsilon$ value that maximizes the frequency deviations. Due to the anharmonicity of interatomic potential at large number of vibrational quanta, and the expected atom-like behavior of the system near the dissociation limit, the sensitivity to $\delta M$ decreases with large $\Delta\upsilon$. Optimal detection of frequency deviations occurs for transition between $\upsilon=0$ and $\upsilon'$ such that the excited state vibrational energy is a fraction of the dissociation energy \cite{PhysRevLett.100.043202}. Selection of such an optimal transition requires sensitivity calculations  that should include the sensitivity of the anharmonicity constant $\chi_{\rm e}$ \cite{FLConstantin}.
The precise dependences  of the energies of a molecule on FC can in principle be computed {\em ab initio} using quantum chemistry techniques. The most complete such calculations are possible for the one-electron molecules, i.e. the molecular hydrogen ions \cite{Schiller2005,Korobov2017a,Alighanbari2020}. However, we do not expect that precise calculations are needed in order to interpret the experimental data, unless one deals with a (undesired) situation of measuring the differential frequency fluctuations between two systems having close sensitivities.

We stress that all molecular energy contributions are proportional to the Rydberg constant. Furthermore, there is an additional sensitivity to the electron mass in all molecular energy contributions discussed above. In total,
\begin{equation}
    \frac{\delta f_{\rm mol}}{f_{\rm mol}}=2{\delta\alpha\over \alpha_0}+{\delta m_{\rm e}\over m_{{\rm e},0}}+
    \frac{1}{f_{\rm mol}}\Big( f_{\rm el}{m_{\rm e,0}\over M_{\rm{0}}}
    -{1\over2} f_{\rm vib}
    - f_{\rm rot}\Big)
    \Big({\delta M\over M_{\rm{0}}}-{\delta m_{\rm e}\over m_{{\rm e},0}} \Big).
    \label{eq:fmolvariation1}
\end{equation}
As we have seen in many cases, Eq. (\ref{eq:fmolvariation1}) approximates to
\begin{equation}
    \frac{\delta f_{\rm mol}}{f_{\rm mol}}\approx
    2{\delta\alpha\over \alpha_0}+
    \Big(1+{f_{\rm vib}\over 2f_{\rm mol}}\Big){\delta m_{\rm e}\over m_{{\rm e},0}}
  - {f_{\rm vib}\over 2f_{\rm mol}}{\delta M\over M_{\rm{0}}}.
    \label{eq:fmolvariation2}
\end{equation}
 This shows that molecules have simultaneous sensitivity to electron mass, nuclear mass and fine structure constant. Thus, they are general detectors of FC oscillations. 
The fractional sensitivity to nuclear mass is determined by the ratio ${f_{\rm vib}/ 2f_{\rm mol}}$. It amounts to 0.06  in the example above. For the well-known iodine transition R56(32-0) at 532\,nm, the difference $\upsilon'-\upsilon=32$ is larger than in the above, so that the ratio increases to approximately 0.084. We note that a more accurate sensitivity estimate can be obtained through consideration of the mass dependences of the Dunham  parameters \cite{Herzberg}, as is done, for example, in \cite{FLConstantin}.

The sensitivity of a molecular system to these constants is present not only if one probes an electronic-vibrational transition (as  discussed above) but also if one probes a purely vibrational transition (without change of electronic state).  In this case, $f_{\rm mol}\simeq f_{\rm vib}$ in Eq.\,(\ref{eq:fmolvariation2}).  The ratio relevant to nuclear mass sensitivity increases sensibly to  close to $-0.5$.
The frequencies of such transitions lie in the infrared spectral range \cite{Note3}.


Let us state the sensitivities of  a comparison between a molecular transition frequency and an optical cavity resonance frequency. 
 The comparison consists in measuring the frequency ratio $f_{\rm mol}/f_{\rm L}$. For simplicity, we assume the realistic case that the two frequencies are nearly equal, denoted by $f_{\rm{mol}}$.  Recalling  the transition in sensitivity to FC oscillations around the acoustic cut-off frequency $f_{\rm{c1}}$ of the optical cavity (see section \ref{sec:section2}), we combine Eq.\,(\ref{eq:fmolvariation2}) with Eq.\,(\ref{eq:deltafL}) to obtain for the fractional variation of $f_{\rm mol}/f_{\rm L}$: 




\begin{equation}
\label{eq:fmolcases}
\frac{\delta (f_{\rm{mol}}-f_{\rm{L}})}{ f_{\rm{mol}}}=\begin{dcases}
   {\delta\alpha\over \alpha_0}+
    {f_{\rm vib}\over 2f_{\rm mol}}{\delta m_{\rm e}\over m_{{\rm e},0}}
  - {f_{\rm vib}\over 2f_{\rm mol}}{\delta M\over M_{\rm{0}}}, & f\leq f_{\rm{c1}} \\
               2{\delta\alpha\over \alpha_0}+
    \Big(1+{f_{\rm vib}\over 2f_{\rm mol}}\Big){\delta m_{\rm e}\over m_{{\rm e},0}}
  - {f_{\rm vib}\over 2f_{\rm mol}}{\delta M\over M_{\rm{0}}}, & f> f_{\rm{c1}}. \\
    \end{dcases}
\end{equation}


\noindent 



\noindent We can further express Eq. \eqref{eq:fmolcases} in a manner analogous to Eq. \eqref{eq:fatcases} to illustrate the sensitivity in detection of the couplings of the DM field to the photon, electron and nucleons: 
\begin{equation}
\label{eq:fmolcases2}
\frac{\delta (f_{\rm{mol}}-f_{\rm{L}})}{ f_{\rm{mol}}}=\begin{dcases}
  \Big[g_{\rm{\gamma}}+\frac{f_{\rm vib}}{2f_{\rm mol}}\frac{g_{\rm{e}}}{m_{\rm{e,0}}}-\frac{f_{\rm vib}}{2f_{\rm mol}}\frac{g_{\rm{n}}}{M_{\rm{0}}}\Big]\frac{\sqrt{2\rho_{\rm DM}}}{m_{\rm{\phi}}}h_{\rm{mol}}(f), & f\leq f_{\rm{c1}} \\
\Big[2g_{\rm{\gamma}}+\Big( 1+\frac{f_{\rm vib}}{2f_{\rm mol}}\Big)\frac{g_{\rm{e}}}{m_{\rm{e,0}}}-\frac{f_{\rm vib}}{2f_{\rm mol}}\frac{g_{\rm{n}}}{M_{\rm{0}}}\Big]\frac{\sqrt{2\rho_{\rm DM}}}{m_{\rm{\phi}}}h_{\rm{mol}}(f), & f> f_{\rm{c1}}.
    \end{dcases}
\end{equation}

\noindent The molecular response function $h_{\rm{mol}}(f)$ is analogous to the function $h_{\rm{at}}(f)$ of Eq. \eqref{eq:fatcases}.

If we instead compare a molecular transition with an atomic transition that satisfies Eq.\,(\ref{eq:deltafatom}), the sensitivity to $\alpha$ is lost. It can be regained if an atomic transition is employed for which there is a strong contribution from relativistic effects \cite{BKB11,PBFS15}.

\section{Conclusions}

We have reviewed the motivations for extending direct searches for fundamental constant variations to the radio-frequency band, and have described the Weekend Relaxion-Search Laboratory, an experiment designed to probe rapid oscillations of the fine structure constant and the electron mass in the 20 kHz-100 MHz range, via atomic spectroscopy. Apparatus upgrades in the original setup, which was used to provide competitive constraints on scalar Dark Matter within scenarios of relaxion Earth-halos, are expected to enable an improved search for rapid oscillations of the constants, with up to $\approx$1000 times higher sensitivity. We proposed an extension of the work to molecular systems, that will allow a search for oscillations in the nuclear mass, as the related sensitivity is not suppressed when considering molecular vibrational levels. The sensitivity analysis presented in the paper is valid for oscillations occurring at all time scales, long or short. \\

\noindent \textbf{Acknowledgements}
We are grateful to R. Ozeri for insightful discussions. This work was supported by the Cluster of Excellence ``Precision Physics, Fundamental Interactions, and Structure of Matter'' (PRISMA+ EXC 2118/1) funded by the German Research Foundation (DFG) within the German Excellence Strategy (Project ID 39083149), by the European Research Council (ERC) under the European Union Horizon 2020 research and innovation program (project Dark-OST, grant agreement No 695405), by the DFG Reinhart Koselleck project and by Internal University Research Funding of Johannes Gutenberg-University Mainz. The work of MGK was supported by the Russian Science Foundation (RSF) grant No 19-12-00157. The work of GP is supported by grants from BSF-NSF, Friedrich Wilhelm Bessel research award, GIF, ISF,
Minerva, Yeda-Sela-SABRA-WRC, and the Segre Research Award.



\section*{References}

\nocite{*}
\bibliographystyle{unsrt}
\bibliography{BIB}


\end{document}